\documentclass[twocolumn,showpacs,amsmath,amssymb,superscriptaddress]{revtex4}
\usepackage{graphicx}
\usepackage{dcolumn}
\usepackage{bm}
\usepackage{amssymb}
\usepackage{amsmath}
\usepackage{epstopdf}
\usepackage{textcomp}
\usepackage{verbatim}
\begin{document}

\title{Realization of random-field dipolar Ising ferromagnetism in a molecular magnet}

\author{Bo Wen}
\affiliation{Department of Physics, City College of New York, CUNY, New York, New York 10031, USA}
\author{P. Subedi}
\affiliation{Department of Physics, New York University, Washington Place, New York, New York 10003, USA}
\author{Lin Bo}
\affiliation{Department of Physics, City College of New York, CUNY, New York, New York 10031, USA}
\author{Y. Yeshurun}
\affiliation{Department of Physics, City College of New York, CUNY, New York, New York 10031, USA}
\affiliation{Department of Physics, New York University, Washington Place, New York, New York 10003, USA}
\affiliation{Department of Physics Bar-Ilan University Ramat-Gan 52900, Israel}
\author{M. P. Sarachik}
\affiliation{Department of Physics, City College of New York, CUNY, New York, New York 10031, USA}
\author{A. D. Kent}
\affiliation{Department of Physics, New York University, Washington Place, New York, New York 10003, USA}
\author{C. Lampropoulos}
\author{G. Christou}
\affiliation{Department of Chemistry, University of Florida, Gainesville, Florida 32611, USA}

\begin{abstract}
The longitudinal magnetic susceptibility of single crystals of the molecular magnet Mn$_{12}$-acetate obeys a Curie-Weiss law, indicating a transition to a ferromagnetic phase due to dipolar interactions. With increasing magnetic field applied transverse to the easy axis, the transition temperature decreases considerably more rapidly than predicted by mean field theory to a $T=0$ quantum critical point.   Our results are consistent with an effective Hamiltonian for a random-field Ising ferromagnet in a transverse field, where the randomness is induced by an external field applied to Mn$_{12}$-acetate crystals that are known to have an intrinsic distribution of locally tilted magnetic easy axes.

\end{abstract}

\maketitle

Dipolar interactions that can lead to long range ferromagnetic order have been explored extensively for decades, both theoretically \cite{PhysRev.70.954} and experimentally \cite{PhysRevLett.65.1064}.  Recent interest in dipolar ferromagnetism has focused on quantum systems, where quantum fluctuations of the spins compete with the dipolar long range order.  Of particular interest are the rare earth LiHoF$_4$ \cite{RosenbaumPRL1996} and similar Y-doped compounds, the only realizations to date of a random-field Ising ferromagnet (RFIFM) \cite{RosenbaumNature2007}.  In these dipolar ferromagnets, a transverse field introduces spin quantum fluctuations and  an approximately linear reduction of the ferromagnetic transition temperature down to absolute zero.

Finite temperature transitions to dipolar ferromagnetism have been demonstrated in several single molecule magnets (SMMs) \cite{PhysRevLett.90.017206,PhysRevLett.93.117202,LuisPRL2005,evangelisti:167202,FernandezPRB2000,garanin:174425}.  In Mn$_{12}$-acetate (henceforth abbreviated as Mn$_{12}$-ac), the first-synthesized and best-studied example of a SMM, a transition to dipolar ferromagnetism was inferred from neutron scattering experiments by Luis \textit{et al.} \cite{PhysRevLett.93.117202}, and confirmed by Monte Carlo simulations \cite{FernandezPRB2000} as well as calculations based on the Mean Field Approximation (MFA) \cite{garanin:174425}.

In this paper, we present measurements of the longitudinal magnetization and susceptibility of Mn$_{12}$-ac that confirm the finite temperature transition.  We report further that with increasing magnetic field  applied transverse to the easy axis, the transition temperature decreases more rapidly than expected within mean field theory to a $T=0$ K quantum critical point.  Our results can be described by an effective Hamiltonian for a RFIFM in a transverse magnetic field, where the randomness derives from a natural distribution of discrete tilts of the molecular magnetic easy axis \cite{PhysRevLett.91.047203, PhysRevLett.90.217204, PhysRevB.70.094429, AndyJLTP2005}.  Mn$_{12}$-ac is a particularly clean model system for the study of RFIFM in which the intrinsic disorder is small, but where due to the nature of the disorder substantial randomness is induced in the longitudinal fields when a transverse field is applied.  Our findings represent an important new realization of RFIFM in SMMs, a \textit{different} class of materials that may serve as an important archetype for the study of dipolar ferromagnetism in quantum systems.

Mn$_{12}$-ac has been modeled as an Ising dipolar system with a double-well potential. Each Mn$_{12}$ molecule behaves as a nanomagnet with spin $S = 10$ oriented along the crystallographic \textit{c} axis due to strong anisotropy $DS_z^2 \approx 60$ K \cite{JonathanPRL1996}. The spins crystallize in a body centered tetragonal lattice and the distance between them is sufficiently large that the intercluster exchange is negligible compared to the dipolar interaction. Hysteretic behavior due to slow relaxation is observed below a blocking temperature $T_B\approx3$ K. The application of transverse magnetic field $H_\perp$ induces quantum tunneling between opposite spin orientations, accelerating the relaxation process towards thermal equilibrium. However, in transverse fields as high as $5$ T used in our experiments, spin reversal by resonant quantum tunneling is extremely slow at low temperature, impeding a direct study of the ordered phase. Our approach is, therefore, to study the magnetic behavior \textit{above} the transition temperature. Specifically, we measure the longitudinal magnetic susceptibility in the presence of $H_\perp$, and deduce the nature of the magnetic interactions from the temperature dependence of the susceptibility.

\begin{figure}[tb]
\centering
\includegraphics[width=1\linewidth]{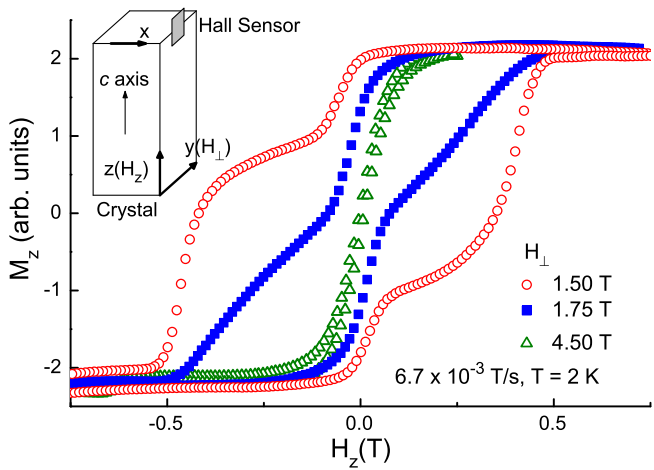}
\caption{(color online). Longitudinal magnetization as a function of the longitudinal field swept at $6.7\times10^{-3}$ T/s for the indicated transverse magnetic fields at $T = 2.0$ K. Inset: Schematic description of the experimental setup.} \label{Hysclose}
\end{figure}

Parallel studies were carried out on the normal Mn$_{12}$-ac, [Mn$_{12}$O$_{12}$(O$_2$CMe)$_{16}$(H$_2$O)$_4$]$\cdot$2MeCO$_2$H$\cdot$4H$_2$O, and a new form [Mn$_{12}$O$_{12}$(O$_2$CMe)$_{16}$(MeOH)$_4$]$\cdot$MeOH, henceforth abbreviated as Mn$_{12}$-ac-MeOH.   The normal form (space group $I\bar{4}$; unit cell parameters $a = b = 17.1668(3)$ \AA, $c = 12.2545(3)$ \AA, $Z = 2, V = 3611.4$ \AA$^3$ at $83$ K) \cite{cornia} and new form (space group $I\bar{4}$; unit cell parameters $a = b = 17.3500(18)$ \AA, $c = 11.9971(17)$, $Z = 2, V = 3611.4$ \AA$^3$ at $173$ K\cite{Stamatatos} of Mn$_{12}$-ac are similar but differ in one crucial aspect of relevance to this work, namely, the solvent molecules of crystallization that lie in-between the Mn$_{12}$ molecules in the crystal. In normal Mn$_{12}$-ac, each Mn$_{12}$ molecule forms O-H..O hydrogen-bonds with $n$ $(n = 0-4)$ of the surrounding MeCO$_2$H molecules, giving a distribution of isomers (of various local symmetries) that ultimately lead to a distribution of easy-axis tilts. In Mn$_{12}$-ac-MeOH, the lattice MeOH molecules form no symmetry-lowering hydrogen bonds to the Mn$_{12}$ molecules, and these crystals have little or no distribution of easy-axis tilts.

Measurements were performed on two Mn$_{12}$-ac single crystals (sample A, dimensions $\sim 0.4 \times 0.4 \times 2.17$ mm$^3$ and sample B, dimensions $\sim 0.4 \times 0.4 \times 2.4$ mm$^3$) and one Mn$_{12}$-ac-MeOH sample (dimensions $\sim 0.2 \times 0.2 \times 0.95$ mm$^3$).  Data for sample B are shown for Mn$_{12}$-ac; sample A displays the same qualitative Êbehavior but has a lower zero transverse field transition temperature ($0.5$K) \cite{sampleA}.  Sample preparation for Mn$_{12}$-ac and Mn$_{12}$-ac-MeOH systems is described in Refs. \cite{Lis} and \cite{Stamatatos}, respectively. A miniature Hall sensor (active area of $50 \times 50\ \mu$m$^2$) was used to measure the magnetization, $M_z$, along the easy direction (\textit{c}-axis) of the crystal.  The sensor was placed near the edge of the sample, where the measured $B_x$ is a linear function of $M_z$.  Care was taken to align the sample and the Hall bar (placed in the y-z plane) relative to each other and relative to the magnet axes.  The point labeled $H_z=0$ was determined by symmetry from full hysteresis loops taken between $-1$ and $1$ T.  Measurements were taken between $0.7$ K and $5.5$ K in a $^3$He refrigerator with a 3D vector superconducting magnet. A longitudinal field, $H_z$, was swept along the sample's easy axis at rates between $1 \times 10^{-5}$ T/s and $6.7 \times 10^{-3}$ T/s, in the presence of a constant transverse field $H_\perp$ (up to 5 T) applied in the $y$ direction (see inset of Fig. \ref{Hysclose}).

The magnetization of Mn$_{12}$-ac is shown in Fig.\ \ref{Hysclose} for different transverse fields at $T = 2.0$ K and longitudinal field sweep rate of $6.7\times10^{-3}$ T/s.  The magnetization exhibits hysteresis due to slow relaxation and the steps characteristic of resonant tunneling \cite{JonathanPRL1996}.  The hysteresis can be eliminated by  applying a transverse field or by sweeping the longitudinal field sufficiently slowly. The effect of transverse field is clearly demonstrated in Fig.\ \ref{Hysclose}: as the transverse field increases, relaxation processes are accelerated, the width of the hysteresis loops decreases, the steps disappear, and equilibrium is ultimately reached so that the magnetization exhibits reversible behavior.

\begin{figure}[b]
\centering
\includegraphics[width=1\linewidth]{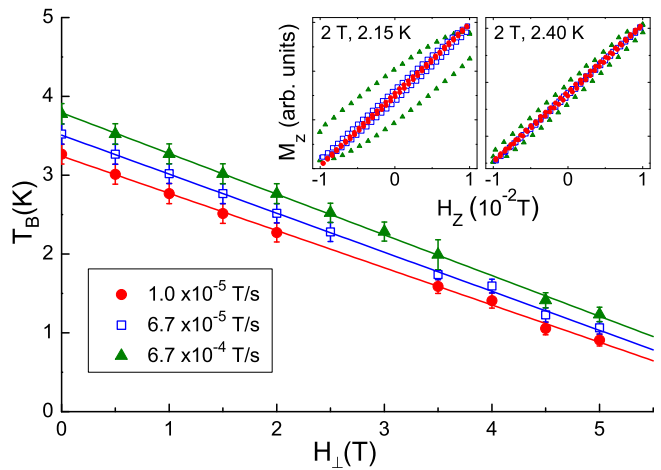}
\caption{(color online). Blocking temperatures for three longitudinal field sweep rates as a function of $H_\perp$. The lines are guides to the eye. Insets: Longitudinal magnetizations as a function of the longitudinal field swept at the indicated rates for $H_\perp$ = 2 T at (a) T $=$ 2.15 K, and (b) T $=$ 2.40 K. } 
\label{dynamicgraph}
\end{figure}

The effect of reducing the sweep rate is demonstrated in the two insets of Fig.\ \ref{dynamicgraph} which show the hysteresis loops obtained for three different sweep rates of longitudinal magnetic field in a narrow range $ \pm 0.01$ T about  $H_z = 0$, measured in the presence of a constant transverse field $H_\perp = 2$ T at $T = 2.15$ K and $T = 2.40$ K.  In each case, hysteresis is observed at the faster sweep rate indicating that the system is below the blocking temperature; at the slower sweep rate the hysteresis loop is closed, indicating the system is above the blocking temperature.  From these and similar data we deduce the field dependence of $T_B$, as summarized in Fig.\ \ref{dynamicgraph} for three different longitudinal-field sweep rates.  $T_B$ was 
estimated by taking the average of the two temperatures where the loops were found to be open and closed at a given sweep rate (see insets to Fig.\ \ref{dynamicgraph}).

The blocking temperature $T_B$ decreases linearly with $H_\perp$, in contrast with single domain uniaxial magnetic particles -- the classical version of Mn$_{12}$ -- where a more moderate decrease is expected of the form $T_B \propto (1-h)^2$, $h = H/H_K$ (here $H$ is the external field and $H_K$ is the anisotropy field \cite{FriedmanPRB1998}).  The faster decrease of $T_B$ in the SMMs suggests that equilibrium is reached more rapidly due to quantum tunneling \cite{FriedmanPRB1998}. 

\begin{figure}[bt]
\centering
\includegraphics[width=1\linewidth]{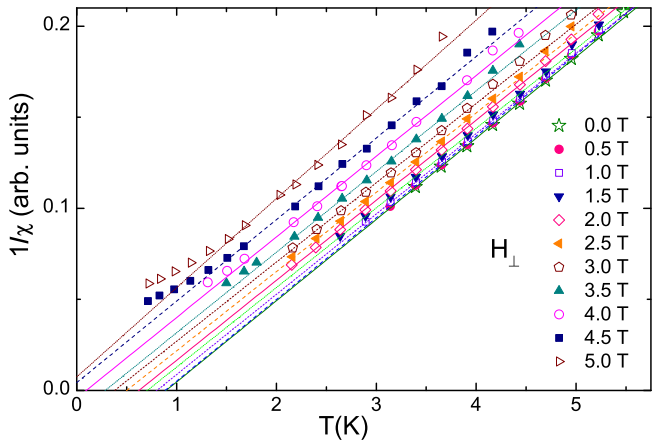}
\caption{(color online). Inverse susceptibility as a function of temperature for different transverse magnetic fields, as indicated. The lines are fits to a Curie-Weiss law (the high field low temperature data are not included in the fit).}
\label{1overXt}
\end{figure}

The longitudinal magnetic susceptibility, $\chi\equiv\partial M_z/\partial H_z|_{H_z=0}$, was deduced from the slope of the reversible $M_z$ versus $H_z$ at $H_z=0$.  Figure \ref{1overXt} shows the inverse of the longitudinal susceptibility as a function of temperature for various transverse fields between zero and 5 T. Except for the high-field data at low temperatures, the susceptibility obeys a Curie-Weiss law, $\chi \propto (T - \theta)^{-1}$, as demonstrated by the straight lines in the figure. As shown by Garanin \cite{Garanin},  the upturn in high transverse field at low temperatures can be attributed to the fact that the tunnel splitting, $\Delta$, becomes larger than $kT$ and, consequently, the susceptibility reflects the quantum state rather than being determined by the temperature. 

\begin{figure}[b]
\centering
\includegraphics[width=1\linewidth]{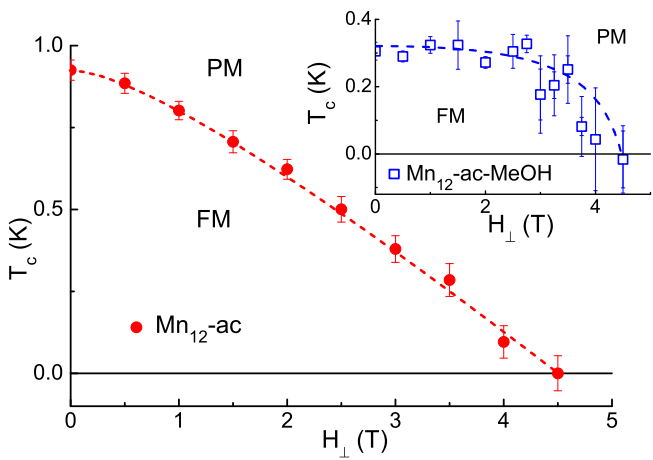}
\caption{(color online). The PM/FM transition temperature for Mn$_{12}$-ac plotted as a 
function of H$_{\perp}$. Inset: The PM/FM transition temperature for Mn$_{12}$-ac-MeOH. The lines are guides to the eye. 
} 
\label{thetaH}
\end{figure}

The temperature intercept $\theta$ deduced for sample B shown in  Fig. \ref{1overXt} depends strongly on $H_\perp$, exhibiting a nearly linear decrease from $\theta \approx 0.90$ K at zero transverse field to zero at approximately $H_\perp \approx 4.5$ T, beyond which negative values are attained. Identifying the Curie-Weiss temperature, $\theta$, as a transition temperature $T_c$ from a paramagnetic (PM) state to an ordered ferromagnetic (FM) phase, one obtains the magnetic phase diagram for Mn$_{12}$-ac of Fig.\ \ref{thetaH}.  In contrast, a similar analysis for Mn$_{12}$-ac-MeOH yields very different behavior, shown in the inset.

The most interesting observation reported in this article is the unexpectedly strong dependence of the transition temperature on transverse field in Mn$_{12}$-ac.  Quantum tunneling of the magnetization in SMMs such as Mn$_{12}$-ac is characterized to lowest order by the spin Hamiltonian  $H=-DS^2_z-g\mu_B\vec{S}\cdot\vec{H}$, where the first term is the uniaxial anisotropy and the second is the Zeeman energy. In zero magnetic field, the up and down states have the same energy. A magnetic field transverse to the anisotropy axis lifts this degeneracy by an energy $\Delta$, the tunnel splitting, leading to quantum mechanical mixing of the Ising up and down spin states at each Mn$_{12}$ site. This introduces channels for quantum relaxation, thereby inducing a decrease in $T_c$. Such a decrease in $T_c$ was observed as a function of transverse field in the rare earth compound LiHo$_{x}$Y$_{1-x}$F$_4$ \cite{RosenbaumPRL1996}. In the undoped material ($x = 1$), $T_c$ is found to decrease as $(1-H_\perp^2)$, consistent with calculations based on MFA. However, the material doped with Y ($x = 0.44$) exhibits behavior similar to the observation reported here, namely an approximately linear decrease of $T_c$ with $H_\perp$. This behavior was attributed to  random-field Ising ferromagnetism (RFIFM), where the random longitudinal fields generated by $H_\perp$ \cite{schechter:137204, RosenbaumNature2007} derive from disorder deliberately introduced by dilution.  In contrast with LiHo$_{x}$Y$_{1-x}$F$_4$, there are no intentional vacancies in Mn$_{12}$-ac. We propose that  Mn$_{12}$-ac provides a clean model system exhibiting RFIFM where a longitudinal  random field is generated by the transverse field due to the presence of  a small amount of $intrinsic$ lattice disorder that is absent in Mn$_{12}$-ac-MeOH.

Disorder in Mn$_{12}$-ac lattice derives from (1) the admixture of a spin-10 faster-relaxing species (at the level of roughly $5$ \%) and (2) as shown in magnetic and EPR studies, the rhombic distortions induced by solvent disorder which results in a distribution of discrete tilts of the molecular magnetic easy axis from the global (average) easy axis of a crystal \cite{PhysRevLett.91.047203,PhysRevLett.90.217204,PhysRevB.70.094429,AndyJLTP2005}.  Neither of these alters the dipolar interactions in a significant way.  However, although the small molecular easy-axis tilts  ($\approx \pm 1^\circ$) induce only a minor perturbation in the dipolar interaction in the absence of field, external transverse fields have projections along (the randomly distributed) easy axes that are comparable in magnitude with the dipolar field itself ($\approx 50$ mT at saturation (see Ref.\ \cite{mchugh:052404}). The application of a transverse field thereby generates substantial randomness in the longitudinal field which depresses $T_c$.   The different behavior we find for Mn$_{12}$-ac, which has a distribution of tilt axes due to isomer disorder, and Mn$_{12}$-ac-MeOH, which has none, is strong evidence that the unusual behavior of the former is indeed due to random fields.

Our observations in Mn$_{12}$-ac can be described by an effective Hamiltonian for interacting Ising spins in a disordered magnet in the presence of transverse field similar to that used by Schechter \cite{schechter:020401} to account for the behavior of the diluted LiHoF$_4$ system:  $H=-\sum_{\langle ij\rangle} J_{ij} S^z_i S^z_j-\Gamma\sum_iS^x_i-\sum_i\gamma_iS^z_i$. Here $S_i$ is the spin at site $i$, $J_{ij}$ is the dipolar interaction between two spins,  $\Gamma$ is the effective transverse field (approximately proportional to the tunnel splitting $\Delta$), and $\gamma_i$ is the effective random field at site $i$.  The effective random field increases approximately linearly with field while $\Gamma$ increases exponentially; in  Mn$_{12}$-ac $\Gamma$ is initially very small and becomes comparable to $\gamma_i$ at about $6$ T.  In the case of Mn$_{12}$-ac-MeOH, the randomness is negligible and the behavior of $T_c$ is controlled by  $\Gamma$, which is quite small in most of the field range of our experiment.   Hence, $T_c$ varies slowly with $H_\perp$ at low fields as shown in the inset of Fig. \ref{thetaH}.  By contrast, in Mn$_{12}$-ac it is the effective random field, $\gamma_i$, that dominates the behavior at low fields, inducing the rapid decrease of $T_c(H_\perp)$ shown in the main panel of the figure.  We note again that Mn$_{12}$-ac provides a clean model system for RFIFM where the relatively small amount of intrinsic disorder introduces only minor deviations from the behavior of the ``perfect'' system in the absence of external field. In contrast with the LiHo$_{x}$Y$_{1-x}$F$_4$ system, where effective transverse fields deriving from off-diagonal terms of the dipolar interaction are introduced by dilution, such terms are excluded by symmetry in Mn$_{12}$-ac.

Confirming the findings of  Luis \textit{et al.} \cite{LuisPRL2005} based on neutron scattering measurements, Mn$_{12}$-ac exhibits a transition to dipolar ferromagnetism at low temperature.   On the other hand, it is puzzling that they found a very different dependence of $T_c$ on transverse field.  We note that there is uncertainty in their determination of $T_c$ from neutron scattering measurements in transverse fields up to 5 T at temperatures below $0.8$ K due to the long time scales required to reach equilibrium (see our Fig.\ \ref{dynamicgraph}).

To summarize, based on measurements of magnetic susceptibility and magnetization, we report that the prototypical single molecule magnet Mn$_{12}$-ac is a new archetype of random-field Ising ferromagnetism in transverse field. In this system, although the intrinsic randomness in the interaction is small, it is sufficient for an externally applied transverse magnetic field to generate a significant random field in the longitudinal direction. The transverse field reduces $T_c$ in two ways: It introduces channels for quantum relaxation for each of the molecules, thereby inducing spin disorder, \textbf{and} it induces fluctuations in the longitudinal dipolar field that are comparable with the intrinsic dipolar interactions themselves, thereby further depressing the ordering temperature. These factors give rise to a dependence of $T_c$ on $H_\perp$ that is inconsistent with  that expected from mean field theory. With increasing transverse field, the transition temperature decreases considerably more rapidly than predicted by mean field theory to a $T=0$ quantum critical point. The critical fluctuations and the behavior approaching the quantum critical point will be discussed elsewhere.

We thank S. McHugh and G. de Loubens for valuable help during the initial phases of the experiment and Zhonghua Zhao for technical assistance.  We acknowledge illuminating discussions with E. Chudnovsky, D. Garanin, A. Millis, A. Mitra, B. Rosenstein, J. R. Friedman, B. Ya. Shapiro, and M. Schechter. Support for GC was provided under grant CHE-0910472; ADK acknowledges support by NSF-DMR-0506946 and ARO W911NF-08-1-0364; MPS acknowledges support from NSF-DMR-0451605; YY acknowledges support of the Deutsche Forschungsgemeinschaft through a DIP project.

\bibliographystyle{apsrev}
\bibliography{rfifm}

\begin{thebibliography}{24}
\expandafter\ifx\csname natexlab\endcsname\relax\def\natexlab#1{#1}\fi
\expandafter\ifx\csname bibnamefont\endcsname\relax
  \def\bibnamefont#1{#1}\fi
\expandafter\ifx\csname bibfnamefont\endcsname\relax
  \def\bibfnamefont#1{#1}\fi
\expandafter\ifx\csname citenamefont\endcsname\relax
  \def\citenamefont#1{#1}\fi
\expandafter\ifx\csname url\endcsname\relax
  \def\url#1{\texttt{#1}}\fi
\expandafter\ifx\csname urlprefix\endcsname\relax\def\urlprefix{URL }\fi
\providecommand{\bibinfo}[2]{#2}
\providecommand{\eprint}[2][]{\url{#2}}

\bibitem[{\citenamefont{Luttinger and Tisza}(1946)}]{PhysRev.70.954}
\bibinfo{author}{\bibfnamefont{J.~M.} \bibnamefont{Luttinger}}
  \bibnamefont{and} \bibinfo{author}{\bibfnamefont{L.}~\bibnamefont{Tisza}},
  \bibinfo{journal}{Phys. Rev.} \textbf{\bibinfo{volume}{70}},
  \bibinfo{pages}{954} (\bibinfo{year}{1946}).

\bibitem[{\citenamefont{Roser and Corruccini}(1990)}]{PhysRevLett.65.1064}
\bibinfo{author}{\bibfnamefont{M.~R.} \bibnamefont{Roser}} \bibnamefont{and}
  \bibinfo{author}{\bibfnamefont{L.~R.} \bibnamefont{Corruccini}},
  \bibinfo{journal}{Phys. Rev. Lett.} \textbf{\bibinfo{volume}{65}},
  \bibinfo{pages}{1064} (\bibinfo{year}{1990}).

\bibitem[{\citenamefont{Bitko et~al.}(1996)\citenamefont{Bitko, Rosenbaum, and
  Aeppli}}]{RosenbaumPRL1996}
\bibinfo{author}{\bibfnamefont{D.}~\bibnamefont{Bitko}},
  \bibinfo{author}{\bibfnamefont{T.~F.} \bibnamefont{Rosenbaum}},
  \bibnamefont{and} \bibinfo{author}{\bibfnamefont{G.}~\bibnamefont{Aeppli}},
  \bibinfo{journal}{Phys. Rev. Lett.} \textbf{\bibinfo{volume}{77}},
  \bibinfo{pages}{940} (\bibinfo{year}{1996}).

\bibitem[{\citenamefont{Silevitch et~al.}(2007)\citenamefont{Silevitch, Bitko,
  Brooke, Ghosh, Aeppli, and Rosenbaum}}]{RosenbaumNature2007}
\bibinfo{author}{\bibfnamefont{D.~M.} \bibnamefont{Silevitch}},
  \bibinfo{author}{\bibfnamefont{D.}~\bibnamefont{Bitko}},
  \bibinfo{author}{\bibfnamefont{J.}~\bibnamefont{Brooke}},
  \bibinfo{author}{\bibfnamefont{S.}~\bibnamefont{Ghosh}},
  \bibinfo{author}{\bibfnamefont{G.}~\bibnamefont{Aeppli}}, \bibnamefont{and}
  \bibinfo{author}{\bibfnamefont{T.~F.} \bibnamefont{Rosenbaum}},
  \bibinfo{journal}{Nature (London)} \textbf{\bibinfo{volume}{448}},
  \bibinfo{pages}{567} (\bibinfo{year}{2007}).

\bibitem[{\citenamefont{Morello et~al.}(2003)\citenamefont{Morello, Mettes,
  Luis, Fern\'andez, Krzystek, Arom\'i, Christou, and
  de~Jongh}}]{PhysRevLett.90.017206}
\bibinfo{author}{\bibfnamefont{A.}~\bibnamefont{Morello}},
  \bibinfo{author}{\bibfnamefont{F.~L.} \bibnamefont{Mettes}},
  \bibinfo{author}{\bibfnamefont{F.}~\bibnamefont{Luis}},
  \bibinfo{author}{\bibfnamefont{J.~F.} \bibnamefont{Fern\'andez}},
  \bibinfo{author}{\bibfnamefont{J.}~\bibnamefont{Krzystek}},
  \bibinfo{author}{\bibfnamefont{G.}~\bibnamefont{Arom\'i}},
  \bibinfo{author}{\bibfnamefont{G.}~\bibnamefont{Christou}}, \bibnamefont{and}
  \bibinfo{author}{\bibfnamefont{L.~J.} \bibnamefont{de~Jongh}},
  \bibinfo{journal}{Phys. Rev. Lett.} \textbf{\bibinfo{volume}{90}},
  \bibinfo{pages}{017206} (\bibinfo{year}{2003}).

\bibitem[{\citenamefont{Evangelisti et~al.}(2004)\citenamefont{Evangelisti,
  Luis, Mettes, Aliaga, Arom\'i, Alonso, Christou, and
  de~Jongh}}]{PhysRevLett.93.117202}
\bibinfo{author}{\bibfnamefont{M.}~\bibnamefont{Evangelisti}},
  \bibinfo{author}{\bibfnamefont{F.}~\bibnamefont{Luis}},
  \bibinfo{author}{\bibfnamefont{F.~L.} \bibnamefont{Mettes}},
  \bibinfo{author}{\bibfnamefont{N.}~\bibnamefont{Aliaga}},
  \bibinfo{author}{\bibfnamefont{G.}~\bibnamefont{Arom\'i}},
  \bibinfo{author}{\bibfnamefont{J.~J.} \bibnamefont{Alonso}},
  \bibinfo{author}{\bibfnamefont{G.}~\bibnamefont{Christou}}, \bibnamefont{and}
  \bibinfo{author}{\bibfnamefont{L.~J.} \bibnamefont{de~Jongh}},
  \bibinfo{journal}{Phys. Rev. Lett.} \textbf{\bibinfo{volume}{93}},
  \bibinfo{pages}{117202} (\bibinfo{year}{2004}).

\bibitem[{\citenamefont{Luis et~al.}(2005)\citenamefont{Luis, Campo, G\'omez,
  McIntyre, Luz\'on, and Ruiz-Molina}}]{LuisPRL2005}
\bibinfo{author}{\bibfnamefont{F.}~\bibnamefont{Luis}},
  \bibinfo{author}{\bibfnamefont{J.}~\bibnamefont{Campo}},
  \bibinfo{author}{\bibfnamefont{J.}~\bibnamefont{G\'omez}},
  \bibinfo{author}{\bibfnamefont{G.~J.} \bibnamefont{McIntyre}},
  \bibinfo{author}{\bibfnamefont{J.}~\bibnamefont{Luz\'on}}, \bibnamefont{and}
  \bibinfo{author}{\bibfnamefont{D.}~\bibnamefont{Ruiz-Molina}},
  \bibinfo{journal}{Phys. Rev. Lett.} \textbf{\bibinfo{volume}{95}},
  \bibinfo{pages}{227202} (\bibinfo{year}{2005}).

\bibitem[{\citenamefont{\relax{M. Evangelisti et
  al.}}(2006)}]{evangelisti:167202}
\bibinfo{author}{\bibnamefont{\relax{M. Evangelisti et al.}}},
  \bibinfo{journal}{Phys. Rev. Lett.} \textbf{\bibinfo{volume}{97}},
  \bibinfo{eid}{167202} (\bibinfo{year}{2006}).

\bibitem[{\citenamefont{Fern\'andez and Alonso}(2000)}]{FernandezPRB2000}
\bibinfo{author}{\bibfnamefont{J.~F.} \bibnamefont{Fern\'andez}}
  \bibnamefont{and} \bibinfo{author}{\bibfnamefont{J.~J.}
  \bibnamefont{Alonso}}, \bibinfo{journal}{Phys. Rev. B}
  \textbf{\bibinfo{volume}{62}}, \bibinfo{pages}{53} (\bibinfo{year}{2000}).

\bibitem[{\citenamefont{Garanin and Chudnovsky}(2008)}]{garanin:174425}
\bibinfo{author}{\bibfnamefont{D.~A.} \bibnamefont{Garanin}} \bibnamefont{and}
  \bibinfo{author}{\bibfnamefont{E.~M.} \bibnamefont{Chudnovsky}},
  \bibinfo{journal}{Phys. Rev. B} \textbf{\bibinfo{volume}{78}},
  \bibinfo{eid}{174425} (\bibinfo{year}{2008}).

\bibitem[{\citenamefont{del Barco et~al.}(2003)\citenamefont{del Barco, Kent,
  Rumberger, Hendrickson, and Christou}}]{PhysRevLett.91.047203}
\bibinfo{author}{\bibfnamefont{E.}~\bibnamefont{del Barco}},
  \bibinfo{author}{\bibfnamefont{A.~D.} \bibnamefont{Kent}},
  \bibinfo{author}{\bibfnamefont{E.~M.} \bibnamefont{Rumberger}},
  \bibinfo{author}{\bibfnamefont{D.~N.} \bibnamefont{Hendrickson}},
  \bibnamefont{and} \bibinfo{author}{\bibfnamefont{G.}~\bibnamefont{Christou}},
  \bibinfo{journal}{Phys. Rev. Lett.} \textbf{\bibinfo{volume}{91}},
  \bibinfo{pages}{047203} (\bibinfo{year}{2003}).

\bibitem[{\citenamefont{Hill et~al.}(2003)\citenamefont{Hill, Edwards, Jones,
  Dalal, and North}}]{PhysRevLett.90.217204}
\bibinfo{author}{\bibfnamefont{S.}~\bibnamefont{Hill}},
  \bibinfo{author}{\bibfnamefont{R.~S.} \bibnamefont{Edwards}},
  \bibinfo{author}{\bibfnamefont{S.~I.} \bibnamefont{Jones}},
  \bibinfo{author}{\bibfnamefont{N.~S.} \bibnamefont{Dalal}}, \bibnamefont{and}
  \bibinfo{author}{\bibfnamefont{J.~M.} \bibnamefont{North}},
  \bibinfo{journal}{Phys. Rev. Lett.} \textbf{\bibinfo{volume}{90}},
  \bibinfo{pages}{217204} (\bibinfo{year}{2003}).

\bibitem[{\citenamefont{Takahashi et~al.}(2004)\citenamefont{Takahashi,
  Edwards, North, Hill, and Dalal}}]{PhysRevB.70.094429}
\bibinfo{author}{\bibfnamefont{S.}~\bibnamefont{Takahashi}},
  \bibinfo{author}{\bibfnamefont{R.~S.} \bibnamefont{Edwards}},
  \bibinfo{author}{\bibfnamefont{J.~M.} \bibnamefont{North}},
  \bibinfo{author}{\bibfnamefont{S.}~\bibnamefont{Hill}}, \bibnamefont{and}
  \bibinfo{author}{\bibfnamefont{N.~S.} \bibnamefont{Dalal}},
  \bibinfo{journal}{Phys. Rev. B} \textbf{\bibinfo{volume}{70}},
  \bibinfo{pages}{094429} (\bibinfo{year}{2004}).

\bibitem[{\citenamefont{del Barco et~al.}(2005)\citenamefont{del Barco, Kent,
  Hill, North, Dalal, Rumberger, Hendrickson, Chakov, and
  Christou}}]{AndyJLTP2005}
\bibinfo{author}{\bibfnamefont{E.}~\bibnamefont{del Barco}},
  \bibinfo{author}{\bibfnamefont{A.~D.} \bibnamefont{Kent}},
  \bibinfo{author}{\bibfnamefont{S.}~\bibnamefont{Hill}},
  \bibinfo{author}{\bibfnamefont{J.~M.} \bibnamefont{North}},
  \bibinfo{author}{\bibfnamefont{N.~S.} \bibnamefont{Dalal}},
  \bibinfo{author}{\bibfnamefont{E.~M.} \bibnamefont{Rumberger}},
  \bibinfo{author}{\bibfnamefont{D.~N.} \bibnamefont{Hendrickson}},
  \bibinfo{author}{\bibfnamefont{N.}~\bibnamefont{Chakov}}, \bibnamefont{and}
  \bibinfo{author}{\bibfnamefont{G.}~\bibnamefont{Christou}},
  \bibinfo{journal}{J. Low Temp. Phys.} \textbf{\bibinfo{volume}{140}},
  \bibinfo{pages}{119} (\bibinfo{year}{2005}).

\bibitem[{Jon()}]{JonathanPRL1996}
\bibinfo{note}{J. R. Friedman, M. P. Sarachik, J. Tejada, and R. Ziolo, Phys.
  Rev. Lett. {\bf 76}, 3830 (1996); L. Thomas, F. Lionti, R. Ballou, D.
  Gatteschi, R. Sessoli, and B. Barbara, Nature (London) {\bf 383}, 145
  (1996).}

\bibitem[{\citenamefont{Cornia et~al.}(2002)\citenamefont{Cornia, Fabretti,
  Sessoli, Sorace, Gatteschi, Barra, Daiguebonne, and Roisnel}}]{cornia}
\bibinfo{author}{\bibfnamefont{A.}~\bibnamefont{Cornia}},
  \bibinfo{author}{\bibfnamefont{A.~C.} \bibnamefont{Fabretti}},
  \bibinfo{author}{\bibfnamefont{R.}~\bibnamefont{Sessoli}},
  \bibinfo{author}{\bibfnamefont{L.}~\bibnamefont{Sorace}},
  \bibinfo{author}{\bibfnamefont{D.}~\bibnamefont{Gatteschi}},
  \bibinfo{author}{\bibfnamefont{A.-L.} \bibnamefont{Barra}},
  \bibinfo{author}{\bibfnamefont{C.}~\bibnamefont{Daiguebonne}},
  \bibnamefont{and} \bibinfo{author}{\bibfnamefont{T.}~\bibnamefont{Roisnel}},
  \bibinfo{journal}{Acta Cryst. Sec. C} \textbf{\bibinfo{volume}{58}},
  \bibinfo{pages}{m371} (\bibinfo{year}{2002}).

\bibitem[{\citenamefont{Stamatatos et~al.}()\citenamefont{Stamatatos, Abboud,
  and Christou}}]{Stamatatos}
\bibinfo{author}{\bibfnamefont{T.~C.} \bibnamefont{Stamatatos}},
  \bibinfo{author}{\bibfnamefont{K.~A.} \bibnamefont{Abboud}},
  \bibnamefont{and} \bibinfo{author}{\bibfnamefont{G.}~\bibnamefont{Christou}},
  \bibinfo{note}{unpublished}.

\bibitem[{sam()}]{sampleA}
\bibinfo{note}{Such differences may be traceable to sample shape and size, to
  different Hall bar locations on the sample, and so on. Studies are underway
  to clarify these issues.}

\bibitem[{\citenamefont{Lis}(1980)}]{Lis}
\bibinfo{author}{\bibfnamefont{T.}~\bibnamefont{Lis}}, \bibinfo{journal}{Acta
  Cryst. B} \textbf{\bibinfo{volume}{36}}, \bibinfo{pages}{2042}
  (\bibinfo{year}{1980}).

\bibitem[{\citenamefont{Friedman}(1998)}]{FriedmanPRB1998}
\bibinfo{author}{\bibfnamefont{J.~R.} \bibnamefont{Friedman}},
  \bibinfo{journal}{Phys. Rev. B} \textbf{\bibinfo{volume}{57}},
  \bibinfo{pages}{10291} (\bibinfo{year}{1998}).

\bibitem[{\citenamefont{Garanin}()}]{Garanin}
\bibinfo{author}{\bibfnamefont{D.~A.} \bibnamefont{Garanin}},
  \bibinfo{note}{unpublished}.

\bibitem[{\citenamefont{Schechter and Laflorencie}(2006)}]{schechter:137204}
\bibinfo{author}{\bibfnamefont{M.}~\bibnamefont{Schechter}} \bibnamefont{and}
  \bibinfo{author}{\bibfnamefont{N.}~\bibnamefont{Laflorencie}},
  \bibinfo{journal}{Phys. Rev. Lett.} \textbf{\bibinfo{volume}{97}},
  \bibinfo{eid}{137204} (\bibinfo{year}{2006}).

\bibitem[{\citenamefont{McHugh et~al.}(2009)\citenamefont{McHugh, Jaafar,
  Sarachik, Myasoedov, Shtrikman, Zeldov, Bagai, and Christou}}]{mchugh:052404}
\bibinfo{author}{\bibfnamefont{S.}~\bibnamefont{McHugh}},
  \bibinfo{author}{\bibfnamefont{R.}~\bibnamefont{Jaafar}},
  \bibinfo{author}{\bibfnamefont{M.~P.} \bibnamefont{Sarachik}},
  \bibinfo{author}{\bibfnamefont{Y.}~\bibnamefont{Myasoedov}},
  \bibinfo{author}{\bibfnamefont{H.}~\bibnamefont{Shtrikman}},
  \bibinfo{author}{\bibfnamefont{E.}~\bibnamefont{Zeldov}},
  \bibinfo{author}{\bibfnamefont{R.}~\bibnamefont{Bagai}}, \bibnamefont{and}
  \bibinfo{author}{\bibfnamefont{G.}~\bibnamefont{Christou}},
  \bibinfo{journal}{Phys. Rev. B} \textbf{\bibinfo{volume}{79}},
  \bibinfo{eid}{052404} (\bibinfo{year}{2009}).

\bibitem[{\citenamefont{Schechter}(2008)}]{schechter:020401}
\bibinfo{author}{\bibfnamefont{M.}~\bibnamefont{Schechter}},
  \bibinfo{journal}{Phys. Rev. B} \textbf{\bibinfo{volume}{77}},
  \bibinfo{eid}{020401 (R)} (\bibinfo{year}{2008}).

\end{thebibliography}

\end{document}